\documentclass[prints]{aa}
\usepackage{graphicx}

\begin{document}
\title{Local galaxy flows within 5 Mpc
\thanks{Based on observations made with the NASA/ESA Hubble Space
Telescope.  The Space Telescope Science Institute is operated by the
Association of Universities for Research in Astronomy, Inc. under NASA
contract NAS 5--26555.}}
\titlerunning{Local galaxy flows }
\author{I. D. Karachentsev \inst{1}
\and D. I. Makarov \inst{1,11}
\and M. E. Sharina \inst{1,11}
\and A. E. Dolphin \inst{2}
\and E. K. Grebel \inst{3}
\and D. Geisler \inst{4}
\and P.~Guhathakurta \inst{5,6}
\and P. W. Hodge \inst{7}
\and V. E. Karachentseva \inst{8}
\and A. Sarajedini \inst{9}
\and P. Seitzer \inst{10}}
\institute{Special Astrophysical Observatory, Russian Academy
of Sciences, N. Arkhyz, KChR, 369167, Russia
\and Kitt Peak National Observatory, National Optical Astronomy
Observatories,
P.O. Box 26732, Tucson, AZ 85726, USA
\and Max-Planck-Institut f\"{u}r Astronomie, K\"{o}nigstuhl 17, D-69117
Heidelberg, Germany
\and Departamento de F\'{\i}sica, Grupo de Astronom\'{\i}a, Universidad de
Concepci\'on,
Casilla 160-C, Concepci\'on, Chile
\and
Herzberg Fellow, Herzberg Institute of Astrophysics, 5071 W.\ Saanich Road,
Victoria, B.C.\ V9E~2E7, Canada
\and Permanent address: UCO/Lick Observatory, University of California at Santa Cruz, Santa
Cruz,
CA 95064, USA
\and Department of Astronomy, University of Washington, Box 351580,
Seattle,
WA 98195, USA
\and Astronomical Observatory of Kiev University, 04053, Observatorna 3,
Kiev,
Ukraine
\and Department of Astronomy, University of Florida, Gainesville, FL
32611,
USA
\and Department of Astronomy, University of Michigan, 830 Dennison
Building,
Ann Arbor, MI 48109, USA
\and Isaac Newton Institute, Chile, SAO Branch}
\date{Accepted: 29/10/2002}

\abstract{
We present Hubble Space Telescope/WFPC2 images of sixteen dwarf galaxies
as part of our snapshot survey of nearby galaxy candidates.  We derive
their distances from the luminosity of the tip of the red giant branch stars
with a typical accuracy of $ \sim 12$ \%. The resulting distances are
4.26 Mpc (KKH 5), 4.74 Mpc (KK 16), 4.72 Mpc (KK 17), 4.66 Mpc (ESO 115-021),
4.43 Mpc (KKH 18), 3.98 Mpc (KK 27), 4.61 Mpc (KKH 34), 4.99 Mpc (KK 54),
4.23 Mpc (ESO 490-017), 4.90 Mpc (FG 202), 5.22 Mpc (UGC 3755), 5.18 Mpc
(UGC 3974), 4.51 Mpc (KK 65), 5.49 Mpc (UGC 4115), 3.78 Mpc (NGC 2915), and
5.27 Mpc (NGC 6503). Based on distances and radial velocities of 156 nearby
galaxies, we plot the local velocity-distance relation, which has a
slope of $H_0 = 73$~km s$^{-1}$ Mpc$^{-1}$ and a radial velocity
dispersion of 85 km s$^{-1}$. When members of the M81 and Cen~A groups
are removed, and distance errors are taken into account, the radial velocity
dispersion drops to $\sigma_v$ = 41 km s$^{-1}$. The local Hubble flow
within 5 Mpc exibits a significant anisotropy, with two infall peculiar
velocity regions directed towards the Supergalactic poles. However, two
observed regions of outflow peculiar velocity, situated on the Supergalactic
equator, are far away ($\sim$ 50$\degr$) from the Virgo/anti-Virgo direction,
which disagrees with a spherically symmetric Virgo-centric flow. About 63\%
of galaxies within 5 Mpc belong to known compact and loose groups. Apart from
them, we found six new probable groups, consisting
entirely of dwarf galaxies.
}
\maketitle

\section{Introduction}
  Until recently very little data have been available to describe
the peculiar velocity field of galaxies around the Local Group (LG).
This surprising situation was caused by the lack of reliable data on
distances (not velocities) for many of the nearest galaxies. The local Hubble
flow has been predicted by  Lynden-Bell (1981) and Sandage (1986) to
be non-linear because of the gravitational deceleration produced by the 
mass of the LG, which could
permit the calculation of the total mass of the LG independently
from mass estimates based on virial motions inside the group. In a larger
volume the deviations from pure Hubble expansion may be
caused by the gravitational action of nearby groups as well as by the
Virgo-centric flow.

   Enormous progress has been made recently in accurate distance
measurements for nearby galaxies beyond the LG  based on the luminosity
of the tip of the red giant branch (TRGB). This method has a precision
comparable to the Cepheid method, but is much faster in terms of
observing time.
Over the last three years, ``snapshot'' surveys of nearby galaxies
using WFPC2 aboard the HST have provided us with distances for about a hundred
nearby galaxies obtained with an accuracy of about 10\% based on
the TRGB method.
Further significant progress is expected in the near future due to
observations with the Advanced Camera for Surveys (ACS) aboard the HST.

  In this paper we present new precise distances to sixteen galaxies
from the general `field' with radial velocities in a range of
160 -- 400 km s$^{-1}$. Together the data on distances to nearby
galaxies published before (Karachentsev et al. 2002a, 2002b, 2002c, 2002d)
as well as data from the literature, this gives us a basis to map the local
field of peculiar velocities for galaxies situated within $\sim$5 Mpc.

\section {WFPC2 photometry and data reduction}

   Images of sixteen galaxies were obtained with the Wide Field and
Planetary Camera (WFPC2) aboard the Hubble Space Telescope (HST) between
October 22, 1999 and July 26 2001 as  part of our HST snapshot survey
of nearby galaxy candidates (Seitzer et al. 1999; Grebel et al. 2000). The galaxies were
observed with 600-second exposures taken in the F606W and F814W filters
for each object.  Digital Sky Survey (DSS) images of them are shown in
Figure 1 with the HST WFPC2 footprints superimposed. The field size
of the red DSS-II images is 6$\arcmin$. Small galaxies were usually
centered on the WF3 chip, but for some bright objects the WFPC2
position was shifted towards the galaxy periphery to decrease
stellar crowding. The WF3 chip images of the galaxies are
presented in upper panels of Figure 2, where both filters are combined.

 For photometric measurements we used the HSTphot stellar photometry
package developed by Dolphin (2000a). The package has been optimized
for the undersampled conditions present in the WFPC2 to work in
crowded fields.  After removing cosmic rays,
simultaneous photometry was performed on the F606W and F814W frames
using \textit{multiphot}, with corrections to an aperture of
radius $0\farcs5$. Charge-transfer efficiency (CTE) corrections and
calibrations were then applied, which are based on the Dolphin (2000b)
formulae, producing $V,I$ photometry for all stars detected in both
images. Additionally, stars with a signal-to-noise ratio $S/N < 3 $,
$\mid \chi \mid\,\, >2.0$, or  $\mid$ sharpness $\mid\,\, >0.4$  in each
exposure were eliminated from the final photometry list. The uncertainty
of the photometric zero point is estimated to be within $0\fm05$
(Dolphin 2000b).

\section{Color-magnitude  diagrams and distances to sixteen nearby galaxies}

The tip of red giant branch (TRGB) method provides an efficient tool
to measure galaxy distances. The TRGB distances agree with those given by
the Cepheid period-luminosity relation to within 5\%. As shown by
Lee et al. (1993), the TRGB is relatively independent
of age and metallicity. In the I band the TRGB for low-mass stars
is found to be stable within $ \sim 0.1$ mag (Salaris \& Cassisi 1997;
Udalski et al. 2001) for metallicities, [Fe/H], encompassing the entire
range from $-$2.1 to $-$0.7 dex found in Galactic globular clusters.
According to Da Costa \& Armandroff (1990), for metal-poor systems the TRGB
is located at $M_I = -4.05$ mag. Ferrarese et al. (2000) calibrated the
zero point of the TRGB from galaxies with Cepheid distances and estimated
$M_I = -4\fm06 \pm0\fm07(random)\pm0.13(systematic)$. A new TRGB calibration,
$M_I = -4\fm04 \pm0\fm12$, was made by Bellazzini et al.(2001) based on
photometry and on a distance estimate from a detached eclipsing binary in the
Galactic globular cluster $ \omega$ Centauri. For this paper we use
 $M_I = -4\fm05$. The lower left panels of Figure 2 show
$I$, $(V-I)$ color-magnitude diagrams (CMDs) for the sixteen
observed galaxies as well as for their surrounding ``field'' regions.

   We determined the TRGB using a Gaussian-smoothed $I$-band
luminosity function (LF) for red stars with colors $V-I$ within $\pm0\fm5$
of the mean $\langle V-I \rangle$ for expected red giant branch stars. Following
Sakai et al. (1996), we applied a Sobel edge-detection filter.
The position of the TRGB was identified with the peak in the
filter response function. The resulting LFs and the Sobel-filtered LFs
are shown in the lower right corners of Figure 2. The results are
summarized in Table 1. There we list:
(1) galaxy name; (2) equatorial coordinates of the
galaxy center; (3,4) apparent integrated magnitude and angular dimension
from the NASA Extragalactic Database (NED); (5) radial velocity with
respect to the LG centroid (Karachentsev \& Makarov 1996);
here we used new accurate velocities measured
by Huchtmeier et al. (2003) for some galaxies; (6) morphological
type in de Vaucouleurs' notation; (7) position of the TRGB and its
uncertainty as derived with the Sobel filter; (8) Galactic extinction
in the  $I$-band (Schlegel et al. 1998); (9) true distance modulus
with its uncertainty, which takes into account the uncertainty in the TRGB,
as well as uncertainties of the HST photometry zero point
($\sim0\fm05$), the aperture corrections ($\sim0\fm05$), and the crowding effects
($\sim0\fm06$) added quadratically; the uncertainties in the extinction and
reddening are taken to be $10\%$ of their values from Schlegel et al.(1998);
[ for more details on the total budget of internal and external systematic errors
for the TRGB method see Mendez et al. (2002)]; and (10) linear distance
in Mpc and its uncertainty. Below, some individual properties of the
galaxies are briefly discussed.

  {\em KKH 5.} This dwarf irregular galaxy of low surface brightness
was discovered by Karachentsev et al. (2001a). It is situated in the Zone of
Avoidance at the periphery of the Maffei/IC342 group. The galaxy appears to
be well resolved into stars. Its CMD (Fig. 2) reveals a sequence
of blue stars with a Galactic foreground extinction of $E(V-I)$ = 0.39 mag
(Schlegel
et al. 1998). The tip of the RGB stars is also seen. The stars above the RGB
are likely to be asymptotic giant branch (AGB)
stars. The CMD for a nearby field of the same area
(the middle panel in the bottom row) shows that the CMD of the galaxy is
not strongly contaminated by foreground stars in spite of its position
at a low galactic latitude, $b = - 11.3\degr$. We determined the TRGB
to be $24\fm65\pm0\fm15$, which corresponds to a distance modulus of
$28\fm15\pm0\fm17$.

  {\em KK 16.} We present the first deep CMD of this dIrr.
Judging by its radial velocity, $V_{LG} = 400$ km/s, KK 16 is a
dwarf companion of the other, brighter dwarf galaxy NGC 784 $(V_{LG} =
386$ km s$^{-1}$). The CMD (Fig. 2) shows a prominent RGB as well as
some blue main-sequence stars and AGB stars. From the TRGB position of
$24\fm46\pm0\fm22$ we obtain a distance modulus of $28\fm38\pm0\fm24$
yielding a linear distance of $4.74\pm0.50$ Mpc. This distance agrees well
with the distance $5.0\pm0.9$ Mpc derived for NGC 784 from its brightest
stars (Drozdovsky \& Karachentsev, 2000).

  {\em KK 17.} Like KK 16, this dwarf irregular galaxy of low surface
brightness is a companion of NGC 784. Its CMD is dominated
by red stars yielding a TRGB magnitude of $24\fm43\pm0\fm17$, the same as
for KK 16 within the errors. The group of three dwarf galaxies, NGC 784,
KK 16, and KK 17 has a radial velocity dispersion of 16 km s$^{-1}$, reminding
of another group of four dwarfs: NGC 3109, Sex A, Sex B, and Antlia at the
edge of the Local group. Tully et al. (2002) consider such loose systems
as groups of `squelched' galaxies in a common dark halo.

 {\em E 115--021 = PGC 09962 = RFGC 566.} Due to its size, $7\farcm2$ by
$0\farcm8$, this edge-on irregular galaxy extends far beyond the WFPC2
field. The CMD shows a large number of AGB stars above the RGB.  The
TRGB is located at $24\fm33\pm0\fm20$. The `field' in Fig. 2
corresponds to the WF2 field away from the main galaxy body. It is
populated mostly with RGB stars having about the same TRGB magnitude
as the body of the galaxy.

 {\em KKH 18.} This is a very isolated dIrr, box-like galaxy. The
CMD shows a mixed population of red and blue stars. The TRGB at
$24\fm57\pm0\fm22$, yields a distance modulus of $28\fm23\pm0\fm24$.

 {\em KK 27 = AM 0319--662.} The object has a smooth regular shape typical
of dwarf spheroidal galaxies. It is located 18$\arcmin$ northeast of the
prominent spiral galaxy NGC 1313, which has $V_{LG} = 270$ km s$^{-1}$. In Fig. 3 of
Ryder et al. (1995)  KK 27 is indicated by an arrow. It was observed in the
HI line but not detected by Huchtmeier et al. (2000). The CMD
appears to be populated mostly by RGB stars with I(TRGB) = $24\fm10\pm0\fm18$,
which yields $(m-M)_0 = 28\fm00\pm0\fm20$. This distance modulus
agrees well with the distance modulus  $28\fm09\pm0\fm06$ derived for
NGC 1313 by Mendez et al. (2002), which confirms that KK 27 is a dSph companion
to NGC 1313.

 {\em KKH 34 = Mailyan 13.} This dIrr galaxy of low surface brightness
with a radial velocity $V_{LG} = 299$ km s$^{-1}$ (Karachentsev et al. 2001a)
is located at the outskirts of the Maffei/IC342 group. It is well resolved
into stars, and its CMD (Fig. 2) shows a mixed population of blue and red
stars. There is no strong discontinuity in the luminosity function but there
is only a slight hint of a red giant branch. Two peaks are seen in the
Sobel--filtered LF. The first peak appears to be caused by AGB stars,
and the second one, at $I = 24\fm75\pm0\fm15$, which we interpret as the TRGB,
yields a distance modulus of $28\fm32\pm0\fm17$.

 {\em KK 54 = ESO 489--056.} This is an isolated dwarf irregular galaxy
with a radial velocity $V_{LG} = 263$ km s$^{-1}$, which is superimposed on a background
spiral galaxy (see Fig. 2). KK 54 is situated at a high Supergalactic
latitude, SGB = $-77.5\degr$. The CMD reveals a mixed population of
blue and red stars. The Sobel--filtered luminosity function shows a
probable peak at $I = 24\fm57$, which corresponds to
a distance modulus of $28\fm49$.

 {\em ESO 490--017 = PGC 19337.} This is a dIrr galaxy with a radial velocity of
268 km s$^{-1}$, which is also situated at a high Supergalactic latitude ($-79.0\degr$).
The galaxy extends over all WFPC2 fields with the brightest part
being centered on the WF3. The CMDs for the central (WF3) and
the peripheric (WF4) regions of PGC 19337 are shown in Fig. 2. In both
fields we find the TRGB to be at $24\fm23\pm0\fm21$, giving a
distance modulus of $28\fm13\pm0\fm23$.

 {\em FG 202 = PGC 20125.} This irregular galaxy of low surface
brightness was found by Feitzinger \& Galinski (1985). It extends far
beyond the WFPC2 field. The CMD is populated by blue and red
stars. The tip of the RGB is seen just above the detection limit at
$I = 24\fm63\pm0\fm20$, which corresponds to a distance modulus of
$28\fm45\pm0\fm22$.

 {\em UGC 3755.} This is a very isolated irregular galaxy at a high
supergalactic latitude ($-63.4\degr$) with a radial velocity $V_{LG} =
190$ km s$^{-1}$. The galaxy was resolved into stars for the first time by
Georgiev et al. (1997) who estimated its distance modulus to be
$28\fm08\pm0\fm40$ from the luminosity of the brightest blue stars.
Recently Mendez et al. (2002) have observed UGC 3755 with the WFPC2 and
determined the TRGB distance modulus to be $28\fm52\pm0\fm07$. From the
derived CMD ( Fig. 2) we found the TRGB position to be
$24\fm71\pm0\fm24$ and a corresponding distance modulus of $28\fm59\pm0\fm25$.
Our exposures of UGC 3755 are likely not long enough to
determine the true magnitude of the TRGB. The presence of many probable
AGB stars makes it difficult to define reliably the tip of the RGB.

 {\em UGC 3974 = DDO 47.} Like UGC 3755, this dIrr galaxy is located
at a high supergalactic latitude ($-55.5\degr$). The galaxy has
a low radial velocity, $V_{LG} = 160$ km s$^{-1}$, and appears to be well
resolved into stars. The CMD shows a mixed population of
red and blue stars with a hint of the TRGB near the limiting magnitude
at $I$(TRGB) = $24\fm58\pm0\fm23$. Thus we derive a distance modulus of
$28\fm57\pm0\fm25$, which is probably a lower limit on the galaxy distance.
Using the magnitudes of the brightest stars, Georgiev et al. (1997)
estimated the distance modulus to be $28\fm15\pm0\fm40$.

 {\em KK 65.} KK 65 is situated 15$\arcmin$ away from UGC 3974, having almost
the same low radial velocity, $V_{LG}$ = 168 km s$^{-1}$. As Fig. 2 shows, this dwarf
irregular galaxy has an arc-like shape resembling that of another nearby
dIrr galaxy DDO 165. We estimated the TRGB magnitude to be $24\fm28\pm
0\fm16$, corresponding to a distance modulus of $28\fm27\pm0\fm18$.
The derived distances to KK 65 and UGC 3974 suggest marginally
that they form a binary system.

 {\em UGC 4115.} This dIrr galaxy with a low radial velocity, $V_{LG} =
210$ km s$^{-1}$, belongs probably to the same loose group of dwarf galaxies
as UGC 3755, UGC 3974, and KK 65 (Tully et al. 2002). The galaxy was
resolved into stars by Georgiev et al. (1997), who estimated its distance
modulus via the brightest stars to be $28\fm61\pm0\fm40$. The CMD
in Fig. 2 shows the blue and red stellar populations with an indication of
TRGB at $24\fm71\pm0\fm21$, which gives a distance modulus $(m-M)_0 =
28\fm70\pm0\fm23$. Quite likely this is only a lower limit of the galaxy
distance.

 {\em NGC 2915.} This very isolated blue compact dwarf (BCD) galaxy with a low
radial velocity, $V_{LG} = 184$ km s$^{-1}$, contains two stellar subsystems: a
high surface - brightness blue core and a red diffuse population. Based on
the luminosity of the brightest stars, Meurer et al. (1994) estimated its
distance as $D = 5.3\pm1.6$ Mpc. According to Bureau et al. (1999) the HI
disk of NGC 2915 extends to 22 optical scalelengths, providing a huge reservoir
for star formation. The galaxy is well resolved into stars in Fig. 2.
Its core, located in the WF3, contains a lot of blue and red stars, but the
peripheric regions, indicated in Fig. 2 as `field', are populated almost
entirely with red stars. We determined the tip of the RGB to be  $I$(TRGB) =
$24\fm37\pm0\fm24$, yielding a distance modulus of $27\fm89\pm0\fm26$. The
derived new linear distance, $D = 3.78\pm0.43$ Mpc, ranks NGC 2915 among
the nearest BCD galaxies together with UGC 4483 (3.21 Mpc),
NGC 6789 (3.60 Mpc), and UGC 6456 (4.34 Mpc).

 {\em NGC 6503.} NGC 6503 is a Sc galaxy located at the edge of the
Local Void. The galaxy was resolved into stars for the first time by
Karachentsev \& Sharina (1997), who derived its distance modulus to be
$28\fm57\pm0\fm40$. Our HST observations were directed to the North-West
edge of NGC 6503, which is less contaminated by blue stars. The left
CMD in Fig. 2 corresponds to the entire WFPC2 field. The right
one shows the stellar population in the halo region only (outer parts of WF2
and WF4). For the halo stars we determined the TRGB position at $I$(TRGB)=
$24\fm62\pm0\fm21$, which yields a distance modulus of $28\fm61\pm0\fm23$.

  \section{ Status of the measured distances in the Local Volume }

Apart from 35 members of the Local Group with distances $D < 1.0$ Mpc,
there are so far 191 galaxies with distance estimates $D < 5.5$ Mpc.
Among them 35 galaxies have no measured radial velocities. The present
sample of data on radial velocities and distances of nearby galaxies
is presented in Table 2. Its columns give: (1) galaxy name, (2)
apparent integrated blue magnitude from the NED or some recent sources
(Makarova, 1999, Parodi et al. 2002), (4) Galactic extinction from
Schlegel et al. (1998), (5) heliocentric radial velocity in km s$^{-1}$ from the NED
or recent measurements by Huchtmeier et al. (2003), (6) radial velocity in the
frame of the Local Group,
(7) galaxy distance with indication of the used method:
``Cep'' -- Cepheids, ``RGB'' -- tip of red giant branch stars,
``SBF'' -- surface brightness fluctuations, ``mem'' --
membership of known nearby groups, ``BS'' -- luminosity
of the brightest stars, and ``TF'' --  Tully-Fisher relation.
The last column gives the reference for the distance.

  Figure 3 shows the distribution of the LV galaxies according to their
distances determined using various distance indicators. The three lower panels correspond
to the most reliable methods giving distances with an accuracy of
$\sim$5 -- 15\%. The same error is probably similar for
the members of some nearby groups (around M81, Cen A, and M83)
with well determined average distances. A characteristic error on distances
estimated via brightest stars or via TF- relation might be
$\sim$(20 -- 30)\%. The two upper panels present distance distributions for 35
galaxies without radial velocities, and also for 32 galaxies with distance
estimates from the Hubble relation $D = V_{LG}/H_0$, for which $H_0 =
73$ km s$^{-1}$ Mpc$^{-1}$ is adopted. As seen from the histograms, the TRGB method is, in
practice, the most efficient method to measure distances within
$\sim$5 Mpc. Besides, 99\% of the TRGB distances have been obtained during the
last three years taking advantage of the superior angular resolution of HST. 
It should be noted, however, that so far the relative number
of the LV galaxies with radial velocities and accurate distance estimates
is 111/223 or only 50\%. The remaining 112 galaxies might be suitable
targets for the next snapshot survey with the Advanced Camera at HST.

\section {Local deviations from the Hubble flow}

  The Hubble relation (radial velocity --- distance) for 156 nearby galaxies
is shown in Figure 4. Here galaxies with accurate distance estimates
(``Cep'', ``RGB'', ``SBF'', and ``mem'') are represented by filled circles,
and galaxies with less reliable distance estimates (``BS'' and ``TF'')
by crosses. In the considered volume there are two massive
groups of galaxies around M81 and Cen A, whose average distances of
$3.73\pm0.04$ Mpc (Karachentsev et al. 2002a), and $3.63\pm0.07$ Mpc
(Karachentsev et al. 2002b) are very similar. Members of these two
groups are shown in Fig.4 as open circles and open squares, respectively.
The solid line corresponds to the Hubble relation with $H_0$ = 73 km s$^{-1}$ Mpc$^{-1}$,
curved at small distances because of the decelerating gravitational action of
the Local Group (Sandage 1986) assuming a total mass of $1.3 ~ 10^{12}M_{\sun}$
(Karachentsev et al. 2002c). The Hubble diagram for the LV galaxies
reveals some important properties.
\begin{enumerate}
 \item The largest deviations from the Hubble regression are seen in the
range of distances between 3.5 and 3.8 Mpc. Their evident reason are the
virial motions of galaxies inside the M81 and Cen A groups. Other nearby
groups, in particular those of M83 and IC342/Maffei, also contribute
to the observed dispersion of radial velocities.

 \item The galaxies situated at the near end of the M81 and Cen groups, in the
distance range 2.5 -- 3.4 Mpc, have radial velocities that are on the average
$\sim$60 km s$^{-1}$ larger than the expected Hubble velocities. In contrast, radial
velocities of galaxies within the distance range of 4.0 -- 4.6 Mpc tend to
have velocities systematically below the Hubble regression line. Such a kind
of ``S''- shaped deviation of radial velocities is typical of the vicinity
of a massive attractor (see, for example, Fig. 1 in Tonry et al. 2000), when
galaxies at the front and at the back of the attractor fall towards its center.
In particular, because of this the galaxies UGC 6456 and NGC 4236
behind the M81 group lie in Fig. 4 much lower than the Hubble regression line.
\end{enumerate}

 As was shown by Karachentsev \& Makarov (1996), the local Hubble flow
on a scale of $\sim$5 Mpc is significantly anisotropic. Based on rough
estimates of distances to 145 galaxies obtained from the luminosity of
their brightest stars, Karachentsev \& Makarov (2001) derived that the local
field of peculiar motions can be described as a tensor of the local Hubble
parameter, $H_{ij}$, which has the main values of ($81\pm3$) : ($62\pm3$) :
$(48\pm5)$ in km s$^{-1}$. The minor axis of the corresponding ellipsoid
is directed towards the polar axis of the Local Supercluster,
and the major axis has an angle of $(29\pm5)\degr$ with
respect to the direction towards the center of the Virgo cluster.
Broadly speaking, the observed anisotropy of velocities corresponds to
a Virgo-centric flow, however, a spherically symmetric
Virgo-centric flow does not fit well the observed peculiar velocity field.

  Our new, more accurate data on galaxy distances given in Table 2 confirm the
presence of an anisotropy of the Hubble flow in the Local Volume. In particular,
Figure 4 shows that isolated galaxies situated at high supergalactic
latitudes (UGC 3755, UGC 3974, UGC 4115, and KK 65) have radial velocities
that are about twice lower than expected with $H_0$ = 73 km s$^{-1}$ Mpc$^{-1}$.

  Figure 5 presents the all-sky distribution of 156 galaxies from Table 2
in Supergalactic coordinates. The galaxies with positive and negative
peculiar velocities with regard to the isotropic Hubble flow ($H_0 =
73$ km s$^{-1}$ Mpc$^{-1}$) are represented by open and filled circles, respectively. The
position of the supergiant elliptical galaxy M87 at the center of the Virgo
cluster (SGL = 102.9$\degr$, SGB = $-2.3\degr$) is indicated with an asterisk.
The observed peculiar velocities of galaxies were smoothed with a spatial 2D-Gaussian
filter with dispersion $\sigma = 25\degr$, and then were plotted in Fig. 5
as a contour map with intervals of 20 km s$^{-1}$. As can be seen, the local peculiar
velocity field is quite symmetric about to the Local Supercluster equator.
The most slowly expanding region of the local Hubble flow with an amplitude
of $-80$ km s$^{-1}$ occupies the southern Supergalactic polar cap
(Monoceros constellation). Another negative peculiar velocity area with a
lower amplitude, $-20$ km s$^{-1}$, corresponds to the northern Supergalactic
cap, also pointing towards the Local Void (Draco constellation). Two regions
of outflow peculiar velocity within the +20 km s$^{-1}$ contours lie just
on the Supergalactic equator in the Centaurus and Pisces constellations.
However, they are located far from the Virgo/anti-Virgo directions,
as would be expected in a spherical Virgo-centric flow.

  The same map of the local field of peculiar velocities is shown in Figure 6
in galactic coordinates. Figure 6 is useful for comparison with the all-sky
contour map of the predicted peculiar velocity field (see Fig. 1 in
Mendez et al. 2002). That map derived from the IRAS galaxy distribution
represents deviations from the pure Hubble flow on the shell corresponding to velocity
$V_{LG}$ = 500 km s$^{-1}$. In general, the observed peculiar velocity map fits the
predicted one, but has a 4--6 times lower amplitude and significantly
different positions of the regions of outflow peculiar velocity.

  \section {Peculiar velocity dispersion}

   According to the results of N-body simulations
(Governato et al. 1997; Klypin et al. 2002), the dispersion of the
peculiar motions of field galaxies and group centers around the mean flow,
$\sigma_v$, contains important information on
galaxy formation and the local density of matter, $\Omega_m$.
Sandage  et al. (1972) and Karachentsev (1971) found a radial velocity
dispersion around the local Hubble flow of $\sim$70 km s$^{-1}$. Such ``cold'' random
motions correspond to  $\Omega_m\sim0.1$. Recent observational data on
galaxies situated within 3 Mpc around the LG yield a surprisingly lower
dispersion, $\sigma_v \sim$25 -- 30 km s$^{-1}$ (Karachentsev et al. 2002c).
The peculiar velocities of the centroids of the nearest groups (Local Group, M81
group, Cen A group, M83 group, CVnI cloud) turn out to be $\sim$25 km s$^{-1}$
as well (Karachentsev et al. 2002a, 2002b, 2002c, 2002d).
The observed quiescence of the local Hubble flow can be considered
(Chernin 2001; Baryshev et al.  2001) as a signature of
a vacuum- dominated universe where the velocity perturbations are
adiabatically decreasing.

  There are several ways of considering $\sigma_v$. The dispersion of radial
velocities in Fig. 4 around an isotropic Hubble flow
yields $\sigma_v$ = 85 km s$^{-1}$, in good agreement with the initial
estimate of Sandage et al. (1972). However, when members of the two
groups around M81 and Cen A with their high random motions are excluded,
$\sigma_v$  decreases to 73 km s$^{-1}$. If one considers the dispersion
around the observed anisotropy of the local Hubble flow, $\sigma_v$
drops to 59 km s$^{-1}$. Here we should remember that the distances of galaxies in
Fig. 4 are determined with a typical relative error of $\sim$15\%. With the mean
galaxy distance $<D> = 3.8$ Mpc and $H_0$ = 73 km s$^{-1}$ Mpc$^{-1}$, the mean distance
error corresponds to an error on the radial velocity $H_0\cdot \sigma_D$ = 42 km s$^{-1}$. Thus,
after quadratic subtraction of this error the mean-square peculiar
velocity of galaxies is reduced to $\sigma_v$ = 41 km s$^{-1}$. The
true value of the random motions of isolated galaxies in the Local Volume
may even be slightly lower because the random motions of galaxies within some
other nearby groups (IC342/Maffei, M83, etc.) were ignored.

  As shown by Karachentsev et al. (2002a, 2002b, 2002c), the total mass-to-
blue luminosity ratios of the LG, M81 group, Cen A group, and M83 group
lie within a range of [30 - 65] $ M_{\sun}/L_{\sun} $. The low $M_T/L_B$
ratio of the nearest groups and also the low velocity dispersion of their
centers, $\sim$25 km s$^{-1}$, correspond to a low mean density of
matter in the Local universe, $\Omega_m\sim0.03 - 0.07 $.

  \section {Concluding remarks}

  A general view of the Local Volume within a radius of 5.5 Mpc is presented
in Figure 7. Its upper panel shows the galaxy distribution projected onto
the Supergalactic plane, and the lower panel corresponds to an edge-on
view. Apart from 156 galaxies with radial velocities known so far (shown
with filled circles), we also plot in Figure 7 35 galaxies without
radial velocity estimates (open circles). All of them are dwarf galaxies
of the dSph and dwarf elliptical (dE) morphological types. In the considered
volume there are six known groups, besides the LG,
whose brightest members: M81, NGC 5128 (=Cen A), M83,
IC 342, NGC 4736, and NGC 253 are shown with asterisks. Altogether,
121 galaxies, or 63\% of their total number inside the shell of
$1.0 < D < 5.5$ Mpc, belong to these compact or loose groups.

  Apart from the well-known groups, where 1 or 2 giant galaxies dominate over
other members, there are also some groups consisting entirely 
of dwarf galaxies. Tully et al. (2002) found four groups of this kind,
the principal members of which are NGC 3109, UGC 8760, UGC 3974, and NGC 784,
respectively. In the Local Volume we found six more such groups.
Their complete list is given in Table 3.
The table columns contain: (1) group member names, where the
brightest galaxy ranks first, (2) number of galaxies in the group,
(3) mean distance to the group, (4) mean projected linear radius of the
group, (5) radial velocity dispersion, (6) absolute B magnitude of the
brightest member, (7) integrated luminosity of the group, (8,9) virial
and orbital (Karachentsev et al. 2002a) mass estimate normalized to the
luminosity unit, (10) crossing  time.

  It follows from the presented data that a typical group of dwarf galaxies
(N = 4 members) is characterized by a median projected radius of $\sim$180
kpc, a median velocity dispersion of only 18 km s$^{-1}$, a median absolute
magnitude of the brightest member of $-15.5$ mag, and a median virial/orbital
mass-to-luminosity ratio of (220-440) $M_{\sun}/L_{\sun}$. Tully et al. (2002)
suggest that these galaxy groups contain a large amount of dark matter
as low mass halos, as expected in a $\Lambda$ CDM cosmology, which
have never hosted significant star formation. The high virial
 mass-to-luminosity ratios favour this idea. However, the typical
crossing time for these groups, 23 Gyr, exceeds largely the age of the
Universe, which means that virial/orbital mass estimates are
fictitious. Altogether, about 13\% of the Local Volume galaxies
belong to these loose associations of dwarf galaxies.

  Together with the usual groups and groups of dwarf galaxies, the Local Volume
contains small empty regions of different sizes, which are completely
devoid of any galaxy. The biggest one is known as the Local Void (Tully
1988). In this respect, a study of the topology of the Local Volume would be
of interest for cosmology (Gottlober et al. 2002).

\acknowledgements
{ We thank the referee, J. Lequeux, for his very useful comments.
Support for this work was provided by NASA through grant GO--08601.01--A
from the Space Telescope Science Institute, which is operated by the
Association of Universities for Research in Astronomy, Inc.,
under NASA contract NAS5--26555.
This work was partially supported by
 RFBR grant 01--02--16001 and DFG-RFBR grant 02--02--04012.
 D.G. gratefully acknowledges support from the Chile {\sl Centro de
Astrof\'\i sica} FONDAP No. 15010003.

 The Digitized Sky Surveys were produced at the Space Telescope
Science Institute under U.S. Government grant NAG W--2166. The
images of these surveys are based on photographic data obtained
using the Oschin Schmidt Telescope on the Palomar Mountain and the UK
Schmidt Telescope. The plates were processed into the present
compressed digital form with permission of these institutions.

 This project made use of the NASA/IPAC Extragalactic Database (NED),
which is operated by the Jet Propulsion Laboratory, Caltech, under
contract with the National Aeronautics and Space Administration.}

{}

\onecolumn
\begin{table}
\caption{New distances to nearby field galaxies}
\begin{tabular}{|lcccllllll|} \hline
Name   &    RA(1950)Dec   & $B_t$  &  a$\times$ b & $V_{LG}$ & T & $I$(TRGB)& $A_I$ & $(m-M)_0$  & $D$   \\
       &  hh mm ss  $\degr\degr^{\prime\prime}\arcsec\arcsec$ & mag  & arcmin & km/s &   &  mag   & mag &  mag    &Mpc  \\
\hline &                  &      &        &      &   &        &     &         &     \\
KKH5   &  010435.0 511025 & 17.1 & 0.6$\times$0.4&  304 & 10&  24.65 & 0.55&  28.15  &4.26 \\
       &                  &      &        &      &   &   0.15 &     &   0.17  &0.32 \\
KK16   &  015230.2 274234 & 16.3 & 0.8$\times$0.3&  400 & 10&  24.46 & 0.13&  28.38  &4.74 \\
       &                  &      &        &      &   &   0.22 &     &   0.24  &0.50 \\
KK17   &  015718.1 283526 & 17.2 & 0.6$\times$0.3&  360 & 10&  24.43 & 0.11&  28.37  &4.72 \\
       &                  &      &        &      &   &   0.17 &     &   0.19  &0.40 \\
E115-021& 023629.0-613324 & 13.34& 7.2$\times$0.8&  337 &  8&  24.33 & 0.05&  28.34  &4.66 \\
P09962 &                  &      &        &      &   &   0.20 &     &   0.22  &0.48 \\
KKH18  &  030000.5 332956 & 16.7 & 0.7$\times$0.4&  375 & 10&  24.57 & 0.39&  28.23  &4.43 \\
       &                  &      &        &      &   &   0.22 &     &   0.24  &0.47 \\
KK27   &  032029.5-663004 & 16.5 & 1.2$\times$0.4&   $-$  & $-$3&  24.10 & 0.15&  28.00  &3.98 \\
       &                  &      &        &      &   &   0.18 &     &   0.20  &0.36 \\
Mai13  &  055323.0 732524 & 17.1 & 0.6$\times$0.5&  299 & 10&  24.75 & 0.48&  28.32  &4.61 \\
KKH34  &                  &      &        &      &   &   0.15 &     &   0.17  &0.35 \\
KK54   &  062416.7-261406 & 15.70& 0.6$\times$0.3&  263 & 10&  24.57 & 0.13&  28.49  &4.99 \\
E489-056&                 &      &        &      &   &   0.25 &     &   0.26  &0.58 \\
E490-017& 063555.0-255718 & 14.01& 1.7$\times$1.3&  268 & 10&  24.23 & 0.15&  28.13  &4.23 \\
P19337  &                 &      &        &      &   &   0.21 &     &   0.23  &0.42 \\
FG202   & 070430.0-582700 & 14.95& 3.5$\times$1.7&  269 & 10&  24.63 & 0.23&  28.45  &4.90 \\
P20125  &                 &      &        &      &   &   0.20 &     &   0.22  &0.45 \\
U3755   & 071106.2 103618 & 14.25& 1.7$\times$1.0&  190 & 10&  24.71 & 0.17&  28.59  &5.22 \\
	&                 &      &        &      &   &   0.24 &     &   0.25  &0.57 \\
U3974   & 073902.9 165507 & 13.71& 3.1$\times$3.0&  160 & 10&  24.58 & 0.06&  28.57  &5.18 \\
DDO47   &                 &      &        &      &   &   0.23 &     &   0.25  &0.57 \\
KK65    & 073940.2 164047 & 15.6 & 0.6$\times$0.3&  168 & 10&  24.28 & 0.06&  28.27  &4.51 \\
	&                 &      &        &      &   &   0.16 &     &   0.18  &0.36 \\
U4115   & 075413.6 143117 & 15.23& 1.8$\times$1.0&  210 & 10&  24.71 & 0.06&  28.70  &5.49 \\
	&                 &      &        &      &   &   0.21 &     &   0.23  &0.56 \\
N2915   & 092630.9-762430 & 13.19& 1.9$\times$1.0&  184 & 10&  24.37 & 0.53&  27.89  &3.78 \\
	&                 &      &        &      &   &   0.24 &     &   0.26  &0.43 \\
N6503   & 174958.7 700926 & 10.74& 7.1$\times$2.4&  301 &  6&  24.62 & 0.06&  28.61  &5.27 \\
			  &      &        &      &   &        &0.21 & & 0.23  &0.53 \\
\hline
\end{tabular}
\end{table}
\begin{table}
\caption{Current census of the Local Volume galaxies with 1.0 $< D <$ 5.5 Mpc}
\begin{tabular}{|lcrrrrrll|} \hline
Name    &  RA (B1950) Dec&   $B_t$ &  $A_b$ &  $V_h$ & $V_{lg}$& $D_{MW}$&     &    Notes            \\
\hline
(1)     &        (2)     &    (3)  &   (4)&   (5)&  (6)&  (7)&  (8) & (9) \\
\hline
E349-031& 000540.9-345124& 15.48& 0.05&  207& 216& 2.9 & bs  & Laustsen \&,1977    \\
N55     & 001238.0-392954&  8.84& 0.06&  129& 111& 1.66& tf  & Puche \&,1988       \\
N59     & 001253.0-214318& 13.12& 0.09&  361& 431& 5.30& sbf*& Jerjen \&,1998      \\
E294-010& 002406.2-420756& 15.60& 0.02&  117&  81& 1.92& rgb & Karachentsev \&,2002c\\
DDO226  & 004035.0-223127& 14.36& 0.07&  357& 408& 4.92& rgb & Grebel \&,2003       \\
N247    & 004439.6-210158&  9.86& 0.08&  160& 215& 2.48& tf  & Puche \&,1988        \\
N253    & 004506.9-253354&  7.92& 0.08&  241& 274& 3.94& rgb & Grebel \&,2003       \\
DDO6    & 004721.0-211718& 15.19& 0.07&  295& 348& 3.34& rgb & Grebel \&,2003       \\
N300    & 005231.8-375712&  8.95& 0.06&  144& 114& 2.15& cep & Freedman \&,1992     \\
KKH5    & 010435.0 511025& 17.1 & 1.22&   39& 304& 4.26& rgb & present paper        \\
U685    & 010442.9 162501& 14.22& 0.25&  155& 349& 4.79& rgb & Maiz-Apellaniz \&,2002\\
N404    & 010639.2 352705& 11.21& 0.25&  -48& 195& 3.06& rgb & Karachentsev \&,2002c \\
E245-05 & 014257.9-435054& 12.73& 0.07&  394& 308& 4.43& rgb & Grebel \&,2003        \\
U1281   & 014639.2 322040& 13.03& 0.20&  157& 367& 5.4 & bs  & Makarova \&,1998b     \\
KK16    & 015230.0 274234& 16.3 & 0.29&  207& 400& 4.74& rgb & present paper         \\
KK17    & 015718.0 283526& 17.20& 0.24&  168& 360& 4.72& rgb & present paper         \\
N784    & 015824.8 283609& 12.16& 0.26&  194& 386& 5.0 & bs  & Drozdovsky \&,2000     \\
Cas1    & 020205.0 684618& 16.38& 4.40&   35& 284& 3.4 & mem*& Maffei group          \\
KKH11   & 022103.7 554709& 16.2 & 2.13&   75& 308& 3.4 & mem*& Maffei group          \\
KKH12   & 022351.3 571550& 17.80& 3.44&   70& 303& 3.4 & mem*& Maffei group          \\
Mafffei1& 023250.7 592616& 13.47& 5.05&   15& 246& 3.4 & mem*& Maffei group          \\
E115-21 & 023629.0-613324& 13.34& 0.11&  513& 337& 4.66& rgb & present paper         \\
Maffei2 & 023807.9 592324& 14.77& 7.19&  -17& 212& 3.4 & mem*& Maffei group          \\
Dw2     & 025019.1 584807& 17.97& 5.13&   94& 316& 3.4 & mem*& Maffei group          \\
MB3     & 025154.1 583935& 19.38& 5.64&   59& 280& 3.4 & mem*& Maffei group          \\
Dw1     & 025306.0 584238& 15.01& 6.34&  112& 333& 3.4 & mem*& Maffei group          \\
KKH18   & 030000.6 332956& 16.7 & 0.86&  216& 375& 4.43& rgb & present paper         \\
N1313   & 031739.0-664042&  9.66& 0.47&  475& 270& 4.15& rgb & Mendez \&,2002        \\
KK35    & 034023.7 674226& 15.7 & 2.50&  105& 320& 3.3 & mem*& IC 342 group          \\
I342    & 034158.6 675626&  9.22& 2.41&   31& 245& 3.28& cep & Saha \&,2002          \\
UA86    & 035500.0 665900& 14.2 & 4.06&   67& 275& 2.6 & bs  & Karachentsev \&,1997a \\
CamA    & 041926.0 724127& 14.85& 0.93&  -47& 164& 3.78& rgb & Karachentsev \&,2002a \\
N1569   & 042604.6 644423& 11.86& 3.02& -104&  88& 2.2 & bs  & Greggio \&,1998       \\
N1560   & 042708.2 714629& 12.16& 0.81&  -36& 171& 3.36& rgb & Karachentsev \&,2002a \\
UA92    & 042722.5 633025& 13.8:& 3.42&  -99&  89& 1.8 & bs  & Karachentsev \&,1997a \\
CamB    & 044802.5 670058& 16.71& 0.94&   77& 266& 3.31& rgb & Karachentsev \&,2002a \\
\end{tabular}
\end{table}
\begin{table}
\begin{tabular}{|lcrrrrrll|} \hline
(1)     &        (2)     &    (3)  &   (4)&   (5)&  (6)&  (7)&  (8) & (9) \\
\hline
N1705   & 045306.2-532627& 12.76& 0.03&  627& 400& 5.10& rgb & Tosi \&,2001          \\
UA105   & 050935.6 623122& 13.9 & 1.35&  111& 279& 3.15& rgb & Karachentsev \&,2002c \\
KKH34   & 055323.1 732524& 17.1 & 1.08&  110& 299& 4.61& rgb & present paper         \\
A0554+07& 055454.2 072915& 19.01& 2.55&  428& 340& 5.5 & bs  & Karachentsev \&,1996  \\
E489-56 & 062416.0-261406& 15.70& 0.28&  492& 263& 4.99& rgb & present paper         \\
E490-17 & 063555.0-255718& 14.01& 0.34&  504& 268& 4.23& rgb & present paper         \\
FG202   & 070428.0-582634& 14.95& 0.51&  554& 269& 4.90& rgb & present paper         \\
U3755   & 071106.2 103631& 14.25& 0.38&  315& 190& 5.22& rgb & present paper         \\
N2366   & 072334.2 691827& 11.68& 0.16&   99& 253& 3.19& rgb & Karachentsev \&,2002a \\
N2403   & 073205.4 654240&  8.82& 0.18&  131& 268& 3.30& cep & Freedman \&, 1988     \\
U3974   & 073902.9 165507& 13.71& 0.14&  270& 160& 5.18& rgb & present paper         \\
KK65    & 073939.4 164048& 15.6 & 0.14&  279& 168& 4.51& rgb*& present paper         \\
U4115   & 075413.0 143131& 15.23& 0.12&  338& 210& 5.49& rgb & present paper         \\
HoII    & 081353.4 705213& 11.09& 0.14&  157& 311& 3.39& rgb & Karachentsev \&,2002a \\
KDG52   & 081843.0 711125& 16.35& 0.09&  113& 268& 3.55& rgb & Karachentsev \&,2002a \\
DDO53   & 082933.0 662101& 14.62& 0.16&   19& 150& 3.56& rgb & Karachentsev \&,2002a \\
U4483   & 083207.0 695657& 15.12& 0.15&  156& 304& 3.21& rgb & Dolphin \&,2001       \\
N2915   & 092630.9-762430& 13.19& 1.19&  460& 184& 3.78& rgb & present paper         \\
HoI     & 093600.0 712447& 13.69& 0.21&  139& 291& 3.84& rgb & Karachentsev \&,2002a \\
N2976   & 094310.0 680843& 10.94& 0.30&    3& 139& 3.56& rgb & Karachentsev \&,2002a \\
BK3N    & 094942.0 691218& 18.89& 0.35&  -40& 101& 4.02& rgb & Karachentsev \&,2002a \\
M81     & 095127.6 691813&  7.69& 0.36&  -35& 107& 3.63& cep & Freedman \&,1994      \\
M82     & 095145.2 695511&  9.06& 0.69&  202& 347& 3.53& rgb & Sakai \&,1999         \\
KDG61   & 095300.0 684947& 15.17& 0.31& -116&  23& 3.60& rgb & Karachentsev \&,2000a \\
A0952+69& 095323.4 693038& 16.8 & 0.37&  100& 243& 3.87& rgb & Karachentsev \&,2002a \\
HoIX    & 095327.9 691653& 14.53& 0.35&   46& 188& 3.7 & mem & M81 group             \\
SexB    & 095723.1 053421& 11.85& 0.14&  301& 111& 1.36& rgb & Karachentsev \&,2002c \\
N3077   & 095921.8 685833& 10.46& 0.29&   13& 153& 3.82& rgb & Karachentsev \&,2002a \\
N3109   & 100049.5-255504& 10.39& 0.29&  403& 110& 1.33& rgb & Karachentsev \&,2002c \\
KDG63   & 100118.0 664753& 15.95& 0.41& -129&   0& 3.50& rgb & Karachentsev \&,2000a \\
U5423   & 100125.1 703627& 14.42& 0.34&  348& 496& 5.3 & bs  & Sharina \&,1999       \\
Antlia  & 100147.0-270521& 16.19& 0.34&  362&  66& 1.32& rgb & Aparicio \&,1997      \\
U5456   & 100440.0 103625& 13.84& 0.18&  544& 377& 3.8:& rgb & Maiz-Apellaniz \&,2002\\
SexA    & 100829.5-042646& 11.86& 0.19&  324&  94& 1.42& rgb & Sakai \&,1996         \\
HIJASS  & 101713.0 685706& 20.  & 0.09&   46& 187& 3.7 & mem & M81 group             \\
\end{tabular}
\end{table}
\begin{table}
\begin{tabular}{|lcrrrrrll|} \hline
(1)     &        (2)     &    (3)  &   (4)&   (5)&  (6)&  (7)&  (8) & (9) \\
\hline
HS117   & 101735.9 712405& 16.5 & 0.49&  -37& 116& 3.7 & mem & M81 group             \\
DDO78   & 102248.0 675440& 15.8 & 0.12&   55& 191& 3.72& rgb*& Karachentsev \&,2000a \\
I2574   & 102441.2 684018& 10.84& 0.16&   57& 197& 4.02& rgb & Karachentsev \&,2002a \\
DDO82   & 102647.0 705233& 13.52& 0.19&   56& 207& 4.00& rgb*& Karachentsev \&,2002a \\
KDG73   & 104928.2 694842& 17.20& 0.08&  116& 263& 3.70& rgb & Karachentsev \&,2002a \\
U6456   & 112435.9 791600& 14.32& 0.16& -103&  89& 4.34& rgb & Mendez \&,2002        \\
U6541   & 113045.9 493052& 14.23& 0.08&  250& 304& 3.89& rgb & Karachentsev \&,2002d \\
N3738   & 113304.4 544758& 11.92& 0.05&  228& 305& 4.90& rgb & Karachentsev \&,2002d \\
N3741   & 113325.2 453343& 14.38& 0.10&  230& 264& 3.03& rgb & Karachentsev \&,2002d \\
KK109   & 114433.5 435659& 17.5 & 0.08&  212& 241& 4.51& rgb & Karachentsev \&,2002d \\
U6817   & 114816.8 390931& 13.45& 0.11&  242& 248& 2.64& rgb & Karachentsev \&,2002c \\
E379-07 & 115210.5-331647& 16.60& 0.32&  640& 363& 5.22& rgb & Karachentsev \&,2002b \\
N4068   & 120129.7 525201& 12.93& 0.09&  210& 290& 5.2 & bs  & Makarova \&,1997      \\
N4163   & 120937.5 362651& 13.66& 0.09&  163& 164& 3.6 & bs  & Tikhonov \&,1998      \\
E321-014& 121113.0-375712& 15.22& 0.40&  613& 337& 3.19& rgb & Karachentsev \&,2002b \\
N4190   & 121113.5 365440& 13.52& 0.13&  230& 234& 3.5 & bs  & Tikhonov \&,1998      \\
U7242   & 121142.2 662212& 14.60& 0.08&   68& 213& 4.3 & mem & N4236 group           \\
DDO113  & 121227.1 362948& 15.70& 0.09&  280& 283& 2.86& rgb & Karachentsev \&,2002c \\
N4214   & 121308.2 363619& 10.24& 0.09&  291& 295& 2.94& rgb & Maiz-Apellaniz \&,2002\\
U7298   & 121400.6 523018& 16.00& 0.10&  173& 255& 4.21& rgb & Karachentsev \&,2002d \\
N4236   & 121421.7 694436& 10.06& 0.06&    0& 160& 4.45& rgb & Karachentsev \&,2002a \\
N4244   & 121459.8 380506& 10.67& 0.09&  243& 255& 4.49& rgb & Karachentsev \&,2002d \\
I3104   & 121545.0-792654& 13.63& 1.70&  430& 171& 2.27& rgb & Karachentsev \&,2002c \\
N4395   & 122320.8 334922& 10.61& 0.07&  320& 315& 4.61& rgb & Karachentsev \&,2002d \\
U7559   & 122437.1 372509& 14.12& 0.06&  218& 231& 4.87& rgb & Karachentsev \&,2002d \\
DDO125  & 122515.4 434613& 12.84& 0.09&  195& 240& 2.54& rgb & Karachentsev \&,2002c \\
N4449   & 122545.1 442215&  9.83& 0.08&  201& 249& 4.21& rgb & Karachentsev \&,2002d \\
U7605   & 122611.0 355940& 14.76& 0.06&  310& 317& 4.43& rgb & Karachentsev \&,2002d \\
UA292   & 123613.3 330229& 16.1 & 0.07&  307& 305& 3.1 & bs  & Makarova \&,1998a     \\
N4605   & 123747.5 615257& 10.89& 0.06&  143& 276& 5.2 & bs  & Karachentsev \&,1994  \\
I3687   & 123950.8 384633& 13.75& 0.09&  358& 385& 4.57& rgb & Karachentsev \&,2002d \\
N4736   & 124832.3 412328&  8.74& 0.08&  309& 353& 4.66& rgb & Karachentsev \&,2002d \\
DDO154  & 125139.3 272510& 14.17& 0.04&  375& 355& 4.3 & bs  & Makarova \&,1998a     \\
GR8     & 125610.9 142914& 14.68& 0.11&  214& 136& 2.10& rgb & Dohm-Palmer \&,1998   \\
KK182   & 130212.8-394854& 16.33& 0.44&  613& 360& 3.6 & mem & CenA group            \\
\end{tabular}
\end{table}
\begin{table}
\begin{tabular}{|lcrrrrrll|} \hline
(1)     &        (2)     &    (3)  &   (4)&   (5)&  (6)&  (7)&  (8) & (9) \\
\hline
N4945   & 130230.9-491212&  9.27& 0.76&  560& 296& 3.6 & mem & CenA group            \\
I4182   & 130329.9 375223& 12.41& 0.06&  320& 356& 4.70& cep & Ferrarese \&,2000     \\
DDO165  & 130439.3 675816& 12.85& 0.10&   31& 196& 4.57& rgb & Karachentsev \&,2002a \\
E269-058& 130738.0-464330& 13.29& 0.46&  402& 142& 3.6 & mem*& CenA group            \\
N5023   & 130957.9 441813& 12.82& 0.08&  407& 476& 5.4 & bs  & Sharina \&,1999       \\
E269-66 & 131015.0-443730& 14.59& 0.40&  784& 528& 3.54& sbf*& Jerjen \&,2000        \\
DDO167  & 131110.8 463504& 15.50& 0.04&  163& 243& 4.19& rgb & Karachentsev \&,2002d \\
U8320   & 131216.6 461101& 13.04& 0.07&  194& 273& 4.33& rgb & Karachentsev \&,2002d \\
KK195   & 131820.5-311605& 18.13& 0.27&  564& 338& 4.6 & mem & M83 group             \\
KK196   & 131850.4-444807& 16.14& 0.36&  741& 490& 3.6 & mem & CenA group            \\
N5102   & 131907.0-362206& 10.28& 0.24&  467& 230& 3.40& rgb & Karachentsev \&,2002b \\
KK200   & 132148.1-304243& 16.67& 0.30&  487& 264& 4.63& rgb & Karachentsev \&,2002b \\
N5128   & 132232.9-424524&  7.84& 0.50&  547& 301& 3.66& rgb & Soria \&,1996         \\
I4247   & 132356.5-300611& 14.4 & 0.27&  415& 195& 4.6 & mem*& M83 group             \\
E324-24 & 132442.0-411318& 12.90& 0.47&  513& 270& 3.73& rgb & Karachentsev \&,2002b \\
N5204   & 132743.8 584032& 11.73& 0.05&  203& 341& 4.65& rgb & Karachentsev \&,2002d \\
U8508   & 132847.1 551002& 13.88& 0.06&   62& 186& 2.56& rgb & Karachentsev \&,2002c \\
N5206   & 133041.0-475342& 11.64& 0.52&  571& 322& 3.6 & mem & CenA group            \\
N5229   & 133158.5 481016& 14.10& 0.08&  363& 460& 5.1 & bs  & Sharina \&,1999       \\
N5238   & 133242.6 515209& 13.55& 0.05&  232& 345& 5.2 & bs  & Makarova \&,1998b     \\
E444-78 & 133342.0-285854& 15.53& 0.23&  573& 363& 4.6 & mem & M83 group             \\
N5236   & 133410.9-293648&  8.20& 0.28&  516& 304& 4.5 & SN  & Schmidt \&,1994       \\
HIPASSa & 133428.7-393836& 16.5 & 0.32&  490& 256& 3.6 & mem*& CenA group            \\
E444-84 & 133432.0-274730& 15.06& 0.30&  587& 380& 4.61& rgb & Karachentsev \&,2002b \\
N5237   & 133440.0-423536& 13.23& 0.41&  370& 131& 3.6 & mem & CenA group            \\
N5253   & 133705.0-312313& 10.87& 0.24&  404& 190& 3.90& cep & Saha \&,1995          \\
I4316   & 133729.0-283830& 14.56& 0.24&  589& 382& 4.41& rgb & Karachentsev \&,2002b \\
U8651   & 133744.2 405931& 14.36& 0.03&  202& 272& 3.01& rgb & Karachentsev \&,2002c \\
N5264   & 133847.0-293942& 12.60& 0.22&  477& 268& 4.53& rgb & Karachentsev \&,2002b \\
E325-11 & 134201.0-413630& 13.99& 0.38&  540& 307& 3.40& rgb & Karachentsev \&,2002b \\
HIPASSc & 134536.7-374308& 16.9 & 0.33&  570& 347& 3.6 & mem*& CenA group            \\
HIPASSb & 134815.3-464511& 17.5 & 0.62&  530& 292& 3.6 & mem*& CenA group            \\
U8760   & 134841.5 381605& 14.64& 0.07&  191& 257& 5.1 & bs  & Makarova \&,1998a     \\
KKH86   & 135202.2 042917& 16.8 & 0.12&  287& 209& 2.61& rgb & Karachentsev \&,2002c \\
U8833   & 135238.0 360456& 15.58& 0.05&  226& 285& 3.19& rgb & Karachentsev \&,2002d \\
\end{tabular}
\end{table}
\begin{table}
\begin{tabular}{|lcrrrrrll|} \hline
(1)     &        (2)     &    (3)  &   (4)&   (5)&  (6)&  (7)&  (8) & (9) \\
\hline
E384-016& 135405.0-350524& 15.11& 0.32&  561& 350& 3.72& sbf*& Jerjen \&,2000        \\
N5408   & 140018.0-410811& 12.21& 0.30&  509& 288& 4.81& rgb & Karachentsev \&,2002b \\
KK230   & 140501.5 351809& 17.9 & 0.06&   62& 126& 1.90& rgb & Grebel \&, 2001       \\
DDO187  & 141338.6 231713& 14.38& 0.10&  152& 172& 2.50& rgb & Aparicio \&,2000      \\
DDO190  & 142248.4 444504& 13.10& 0.05&  150& 263& 2.79& rgb & Karachentsev \&,2002c \\
P51659  & 142448.4-460441& 16.50& 0.56&  386& 171& 3.58& rgb & Karachentsev \&,2002b \\
KKR25   & 161237.3 542946& 16.45& 0.04& -139&  68& 1.86& rgb & Karachentsev \&,2001b \\
N6503   & 174958.7 700926& 10.74& 0.14&   43& 301& 5.27& rgb & present paper         \\
N6789   & 191617.0 635254& 13.76& 0.30& -141& 144& 3.60& rgb & Drozdovsky \&,2001a   \\
SagDIG  & 192705.4-174659& 14.12& 0.52&  -77&  23& 1.04& rgb & Karachentsev \&,2002c \\
I5152   & 215926.6-513214& 11.06& 0.11&  124&  75& 2.07& rgb & Karachentsev \&,2002c \\
UA438   & 232347.3-323957& 13.86& 0.06&   62&  99& 2.23& rgb & Karachentsev \&,2002c \\
UA442   & 234109.0-321412& 13.58& 0.07&  267& 299& 4.27& rgb & Grebel \&,2003        \\
KKH98   & 234303.9 382624& 16.7 & 0.53& -137& 151& 2.45& rgb & Karachentsev \&,2002c \\
N7793   & 235515.0-325206&  9.70& 0.08&  229& 252& 3.91& rgb & Grebel \&,2003        \\
\hline
&\multicolumn{8}{l}{\bf Notes:} \\
\multicolumn{1}{l}{NGC 59.}&
\multicolumn{8}{p{10cm}}{The SBF distance from Jerjen et al.(1998) with a zero point correction +0.9 Mpc.}\\
\multicolumn{1}{l}{Cas1.}&
\multicolumn{8}{p{10cm}}{For Cas1 and other heavily obscured Maffei/IC342 group members we
	  adopt the distance 3.4 Mpc obtained as the average RGB distance
	  for CamA, N1560, CamB and UA105, which are less obscured.}\\
\multicolumn{1}{l}{KKH11.}&
\multicolumn{8}{l}{NED gives $V_{LG}$ instead of $V_h$.}\\
\multicolumn{1}{l}{Maffei2.}&
\multicolumn{8}{l}{ $A_b$ from star photometry, not from Schlegel et al. (1998).} \\
\multicolumn{1}{l}{KK35.}&
\multicolumn{8}{l}{$V_h$ from HI (Huchtmeier \& Karachentsev, 2002).}\\
\multicolumn{1}{l}{KK65.}&
\multicolumn{8}{l}{NED gives an incorrect Vh=407.}\\
\multicolumn{1}{l}{DDO78.}&
\multicolumn{8}{l}{$V_h$ for its globular cluster. NED gives an incorrect $V_h=2550.$}\\
\multicolumn{1}{l}{DDO82.}&
\multicolumn{8}{l}{$V_h=180$ in NED is incorrect.}\\
\multicolumn{1}{l}{E269-58.} &
\multicolumn{8}{l}{$V_h$ from HIPASS, in NED $V_h=1853\pm32.$}\\
\multicolumn{1}{l}{E269-66.}&
\multicolumn{8}{l}{ SBF distance from Jerjen et al.(2000) with a zero point correction $-$0.5 Mpc.}\\
\multicolumn{1}{l}{I4247.}&
\multicolumn{8}{p{10cm}}{$V_h$ from HIPASS, in NED $V_h=274\pm65$.
 HIPASSa. New accurate coordinates are given for HIPASSa, as well HIPASSb
	  and HIPASSc.}\\
\multicolumn{1}{l}{E384-16.}&
\multicolumn{8}{p{10cm}}{The SBF distance from Jerjen et al.(2000) with a zero point correction $-$0.5 Mpc.}\\
\end{tabular}
\end{table}

\begin{table}
\caption{ Properties of nearby groups of dwarf galaxies }
\begin{tabular}{|lrrrrrrrrr|} \hline
   Group    &  N&$< D >$&$< R_p>$&$\sigma_V$&$M_1$&$L_B$&$ M_{vir}/L_B$& $M_{orb}/L_B$& $T_{cross}$ \\
\hline
	    &   &   Mpc & kpc &  km/s &  mag &$10^8L_{\sun}$& $M_{\sun}/L_{\sun}$& $M_{\sun}/L_{\sun}$ &  Gyr \\
\hline
N3109, SexB,  & 4 & 1.36 &  414  &  18  & $-$15.57 &  3.58  &  214  &   201 &     23  \\
Antlia, SexA  &   &      &       &      &        &        &       &       &         \\
	      &   &      &       &      &        &        &       &       &         \\
U8760, U8651, & 3 & 3.20 &  162  &   7  & $-$13.23 &  0.59  &  398  &   430 &     23  \\
U8833         &   &      &       &      &        &        &       &       &         \\
	      &   &      &       &      &        &        &       &       &         \\
U8320, U8215, & 4 & 4.20 &   84  &  37  & $-$15.46 &  2.58  &  869  &   948 &      2.3\\
U8308, U8331  &   &      &       &      &        &        &       &       &         \\
	      &   &      &       &      &        &        &       &       &         \\
N4395, N4244, & 5 & 4.43 &  320  &  54  & $-$17.69 & 35.9   &  625  &   452 &      5.9\\
U7559, U7605, &   &      &       &      &        &        &       &       &         \\
IC3687        &   &      &       &      &        &        &       &       &         \\
	      &   &      &       &      &        &        &       &       &         \\
N784, U1281,  & 4 & 4.96 &  184  &  16  & $-$16.58 &  8.52  &   45  &    84 &     12  \\
KK16, KK17    &   &      &       &      &        &        &       &       &         \\
	      &   &      &       &      &        &        &       &       &         \\
U3974, U3755, & 4 & 5.10 &  412  &  19  & $-$14.97 &  3.43  &  222  &  1945 &     22  \\
KK65, U4115   &   &      &       &      &        &        &       &       &         \\
	      &   &      &       &      &        &        &       &       &         \\
Orion, KK49,  & 3 & 5.95 &  300  &  41  & $-$16.33 &  6.94  & 2045  &  2999 &      7.3\\
U3817         &   &      &       &      &        &        &       &       &         \\
	      &   &      &       &      &        &        &       &       &         \\
U3966,        & 2 & 6.25 &  142  &   1  & $-$14.80 &  1.94  &   $-$   &     7 &    142  \\
U3860         &   &      &       &      &        &        &       &       &         \\
	      &   &      &       &      &        &        &       &       &         \\
U5272, KK78,  & 4 & 7.10 &  114  &  14  & $-$14.91 &  1.91  &   33  &   859 &      8.1\\
KKH54, U5186  &   &      &       &      &        &        &       &       &         \\
	      &   &      &       &      &        &        &       &       &         \\
N2337, U3698, & 3 & 7.90 &  174  &   6  & $-$16.77 &  9.09  &    7  &     3 &     27  \\
U3817         &   &      &       &      &        &        &       &       &         \\
\hline                                                                              \\
 Median       & 4 & 5.0  &  179  &  18  & $-$15.52 &  3.5   &  218  &   441 &     23  \\
\hline
\end{tabular}
\end{table}

\clearpage
\begin{figure*}

\caption{ Digital Sky Survey images of 16 nearby field galaxies.
The field size is 6$\arcmin$, North is to the top and East is to the left.
The HST WFPC2 footprints are superimposed.}
\end{figure*}

\begin{figure*}
\caption{{\bf Top}: WFPC2 images of 18 galaxies: KKH 5, KK 16, KK 17,
ESO 115-021, KKH 18, KK 27, KKH 34, KK 54, ESO 490-017, FG 202, UGC 3755,
KK 65, UGC 4115, NGC 2915, and NGC 6503, produced by
combining the two 600s exposures obtained through the F606W
and F814W filters. The arrows point to the North and the East.
{\bf Bottom left}: The color-magnitude diagrams from the WFPC2 data for
the 16 field galaxies.
{\bf Bottom right}: the Gaussian-smoothed $I$-band luminosity function
restricted to red stars (top), and the output of an edge-detection filter
applied to the luminosity function for the 16 galaxies studied here.}
\end{figure*}

\clearpage
\begin{figure*}
\centering
\vspace{5mm}
\includegraphics[width=16cm]{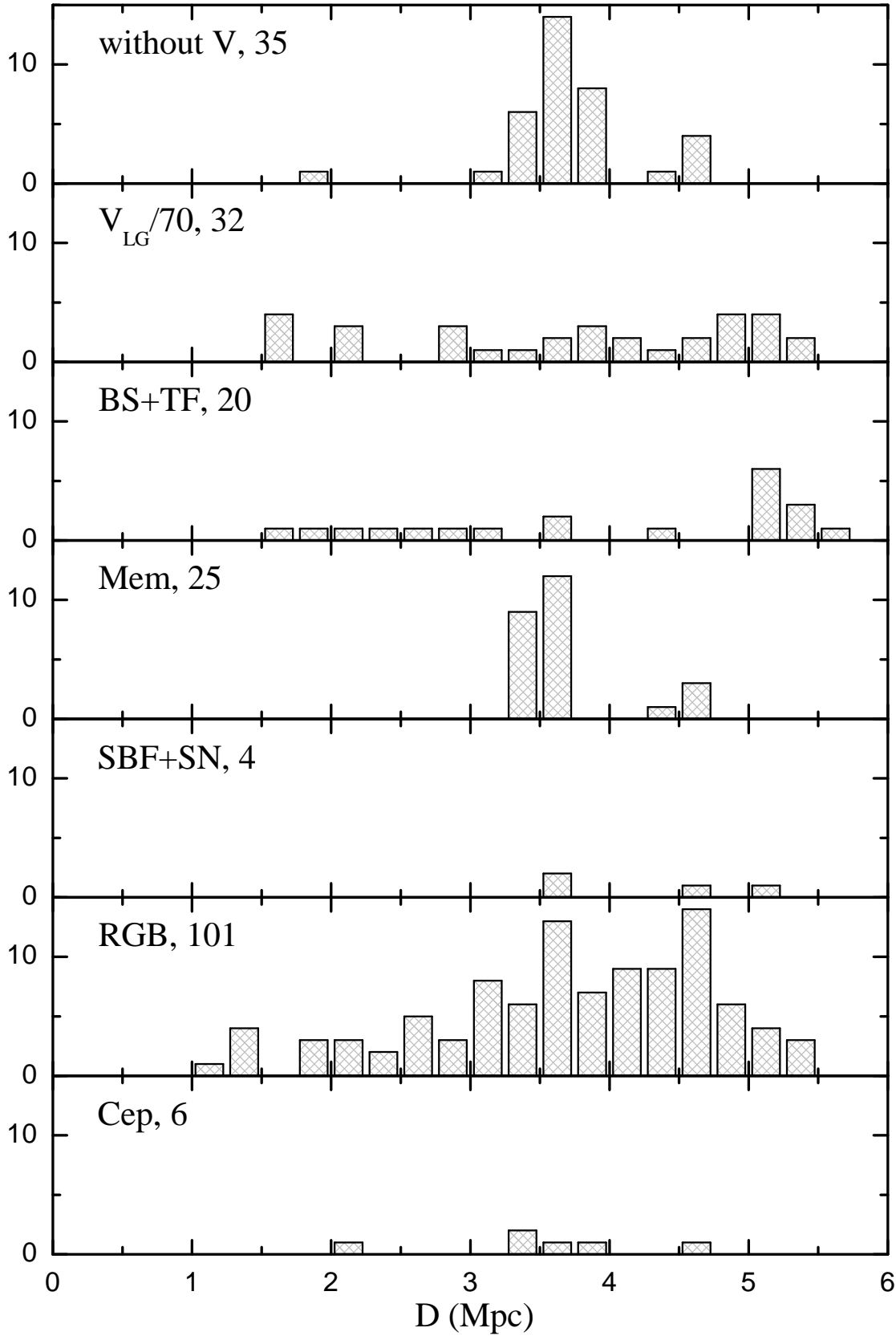}
\caption{ Distribution of 223 Local Volume galaxies according to their
distances derived by different methods:
``Cep'' -- from cepheids, ``RGB'' -- from the tip of red giant branch stars,
``SBF'' -- from surface brightness fluctuations, ``mem'' -- from the galaxy
membership in the known nearby groups, ``BS'' -- from the luminosity
of the brightest stars, and ``TF'' -- from the Tully-Fisher relation.
Two upper panels present distance distributions for 35 galaxies
without radial velocities, as well as for 30 galaxies with distance
estimates from the Hubble relation $D = V_{LG}/H_0$, where $H_0$ =
73 km s$^{-1}$ Mpc$^{-1}$ is adopted.}
\end{figure*}

\begin{figure*}
\centering
\vspace{5mm}

\includegraphics[width=18cm]{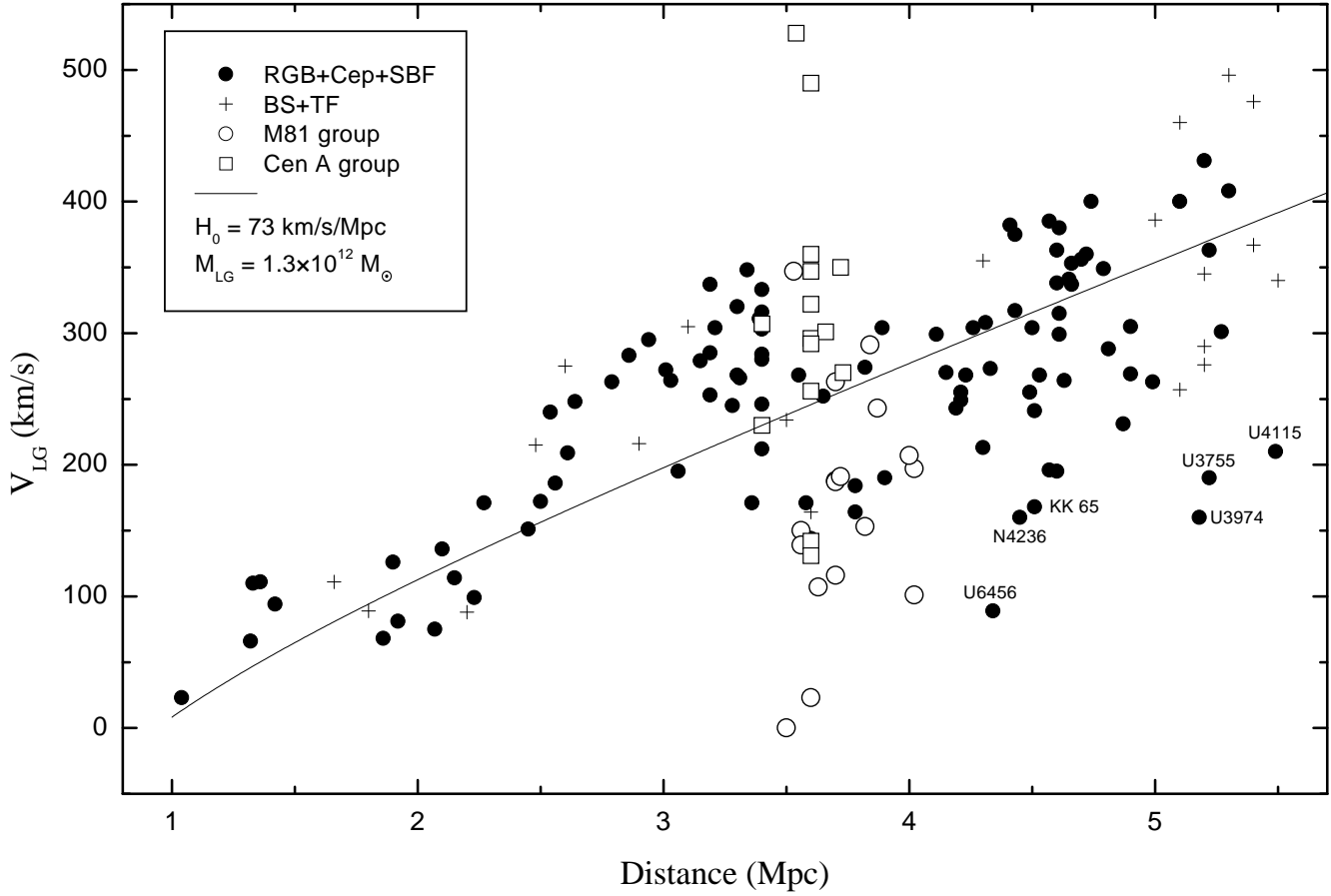}
\vspace{5mm}
\caption{ Radial velocity --- distance relation for 156 Local Volume galaxies.
The galaxies with accurate distance estimates
(``Cep'', ``RGB'', ``SBF'', and ``mem'') are shown as filled circles,
and galaxies with less reliable distance estimates (``BS'' and ``TF'') are
indicated as crosses. The members of M81 and Cen A groups with distances
in the range of 3.4 -- 4.0 Mpc are shown by open circles and open squares,
respectively. The regression line corresponds to the Hubble relation with
$H_0$ = 73 km s$^{-1}$ Mpc$^{-1}$, curved at small distances assuming a decelerating
gravitational action of the Local Group with a total mass of $1.3 ~ 10^{12} M_{\sun}$.}
\end{figure*}

\clearpage
\begin{figure}
\includegraphics[width=18cm]{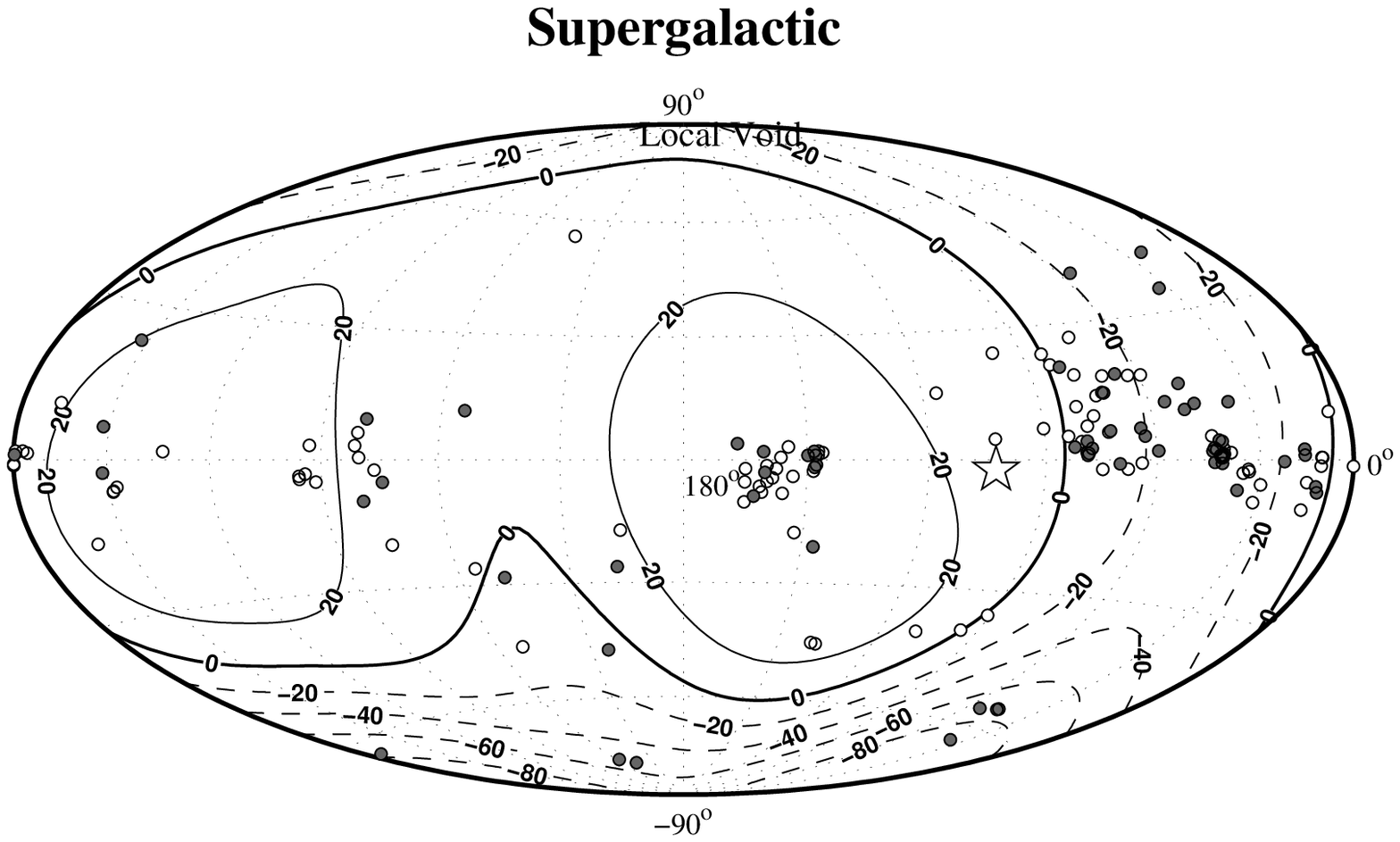}
\vspace{5mm}
\caption{ Full-sky distribution of 156 galaxies from Table 2 in
supergalactic coordinates. The galaxies with positive and with negative
peculiar velocities with respect to the isotropic Hubble flow $(H_0 =
73$ km s$^{-1}$ Mpc$^{-1}$) are shown as open and filled circles, respectively.
The observed peculiar velocities of galaxies were smoothed with a 2D-Gaussian
filter with a parameter $\sigma = 25\degr$, and then were plotted as
contour map with intervals of 20 km s$^{-1}$.}
\end{figure}

\begin{figure*}
\centering
\vspace{5mm}

\includegraphics[width=18cm]{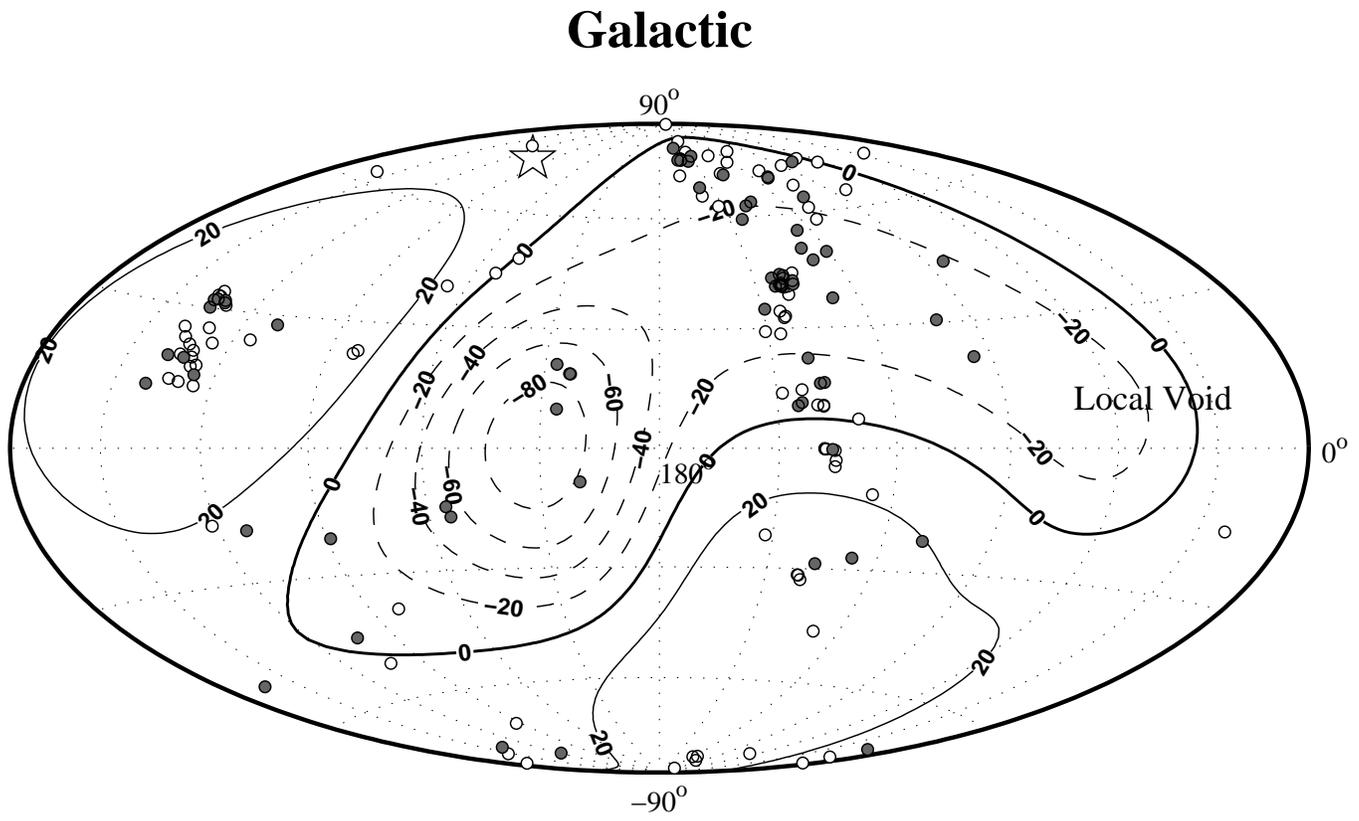}
\vspace{5mm}
\caption{ The same map of the local field of peculiar velocities as shown
in Fig. 5, but in galactic coordinates.}
\end{figure*}

\begin{figure*}
\centering
\vspace{5mm}

\includegraphics[width=18cm]{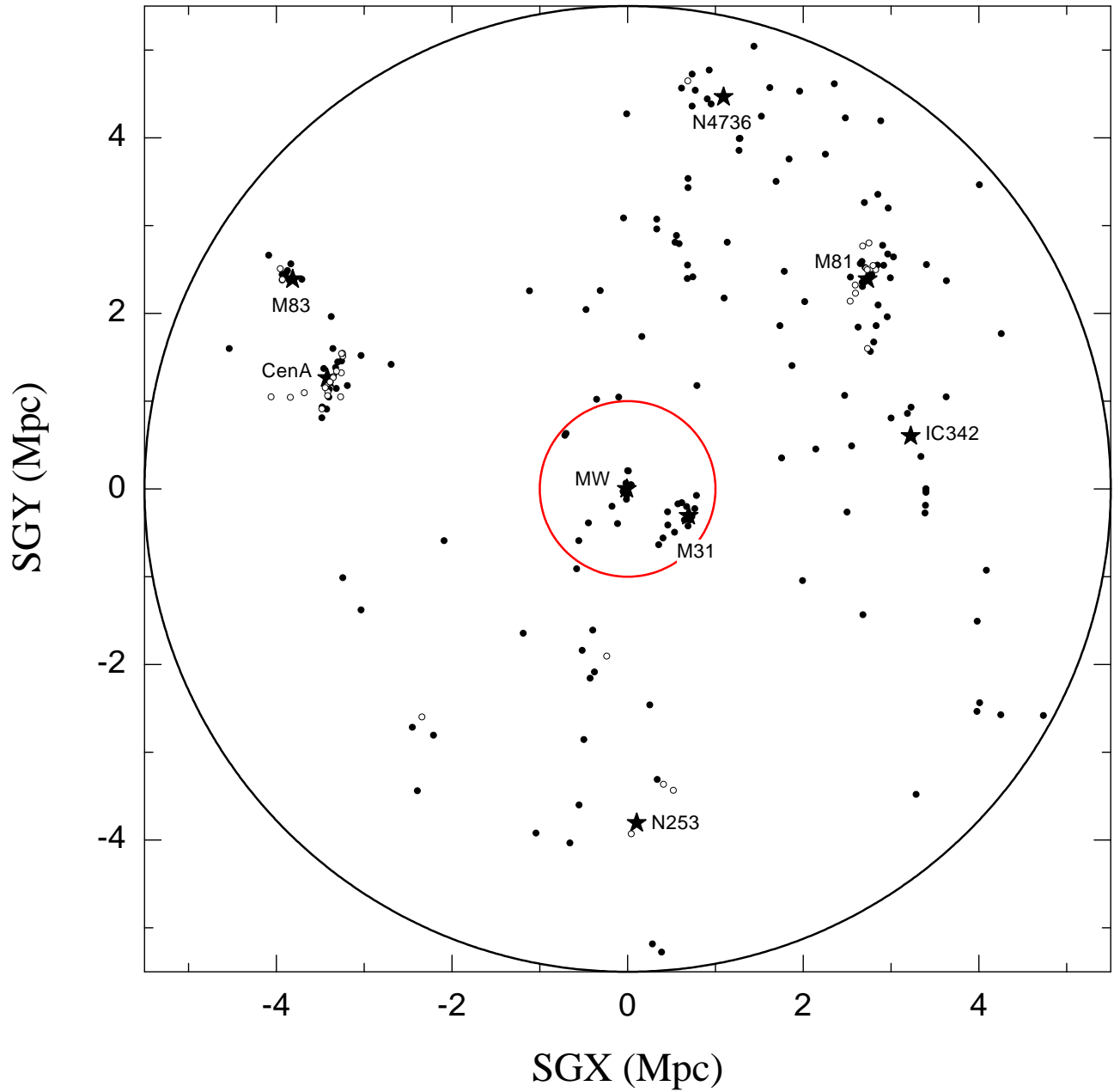}
\vspace{5mm}
\caption{ Panorama of the Local Volume within a radius of 5.5 Mpc.
The upper panel shows the galaxy distribution projected onto the
Supergalactic plane, and the lower panel corresponds to the edge-on
view. The galaxies with known radial velocities are shown as
filled circles, the 35 galaxies of dSph, dE types without radial
velocities are indicated as open circles. The brightest members of
nearest groups are shown as asterisks.}

\end{figure*}

\clearpage
\begin{figure*}
\centering
\vspace{-5mm}
\includegraphics[width=18.0cm]{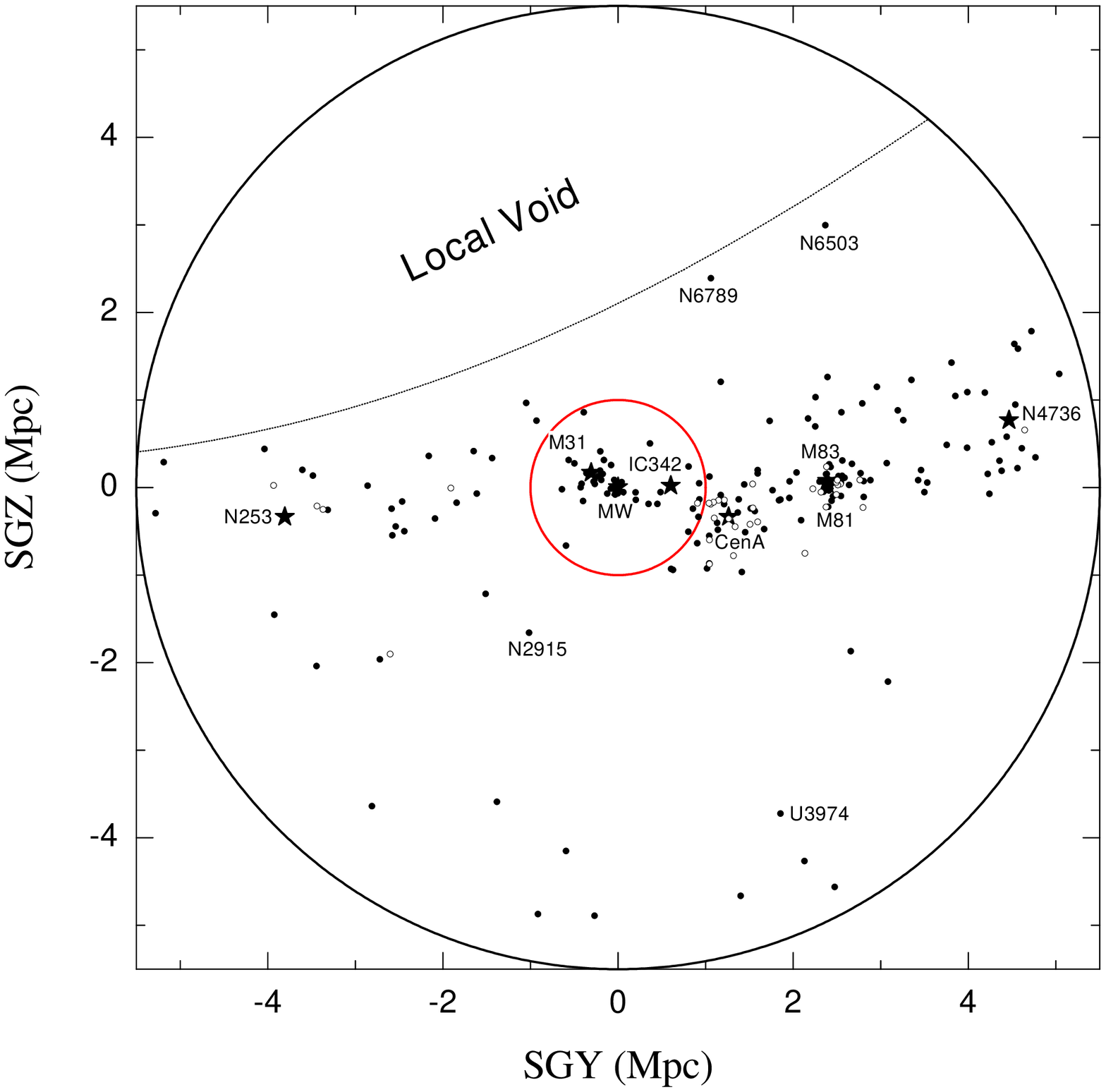}
\setcounter{figure}{6}
\vspace{-5mm}
\caption{continued}
\end{figure*}

\clearpage
\begin{figure*}
\centering
\vspace{-5mm}
\includegraphics[width=18.0cm]{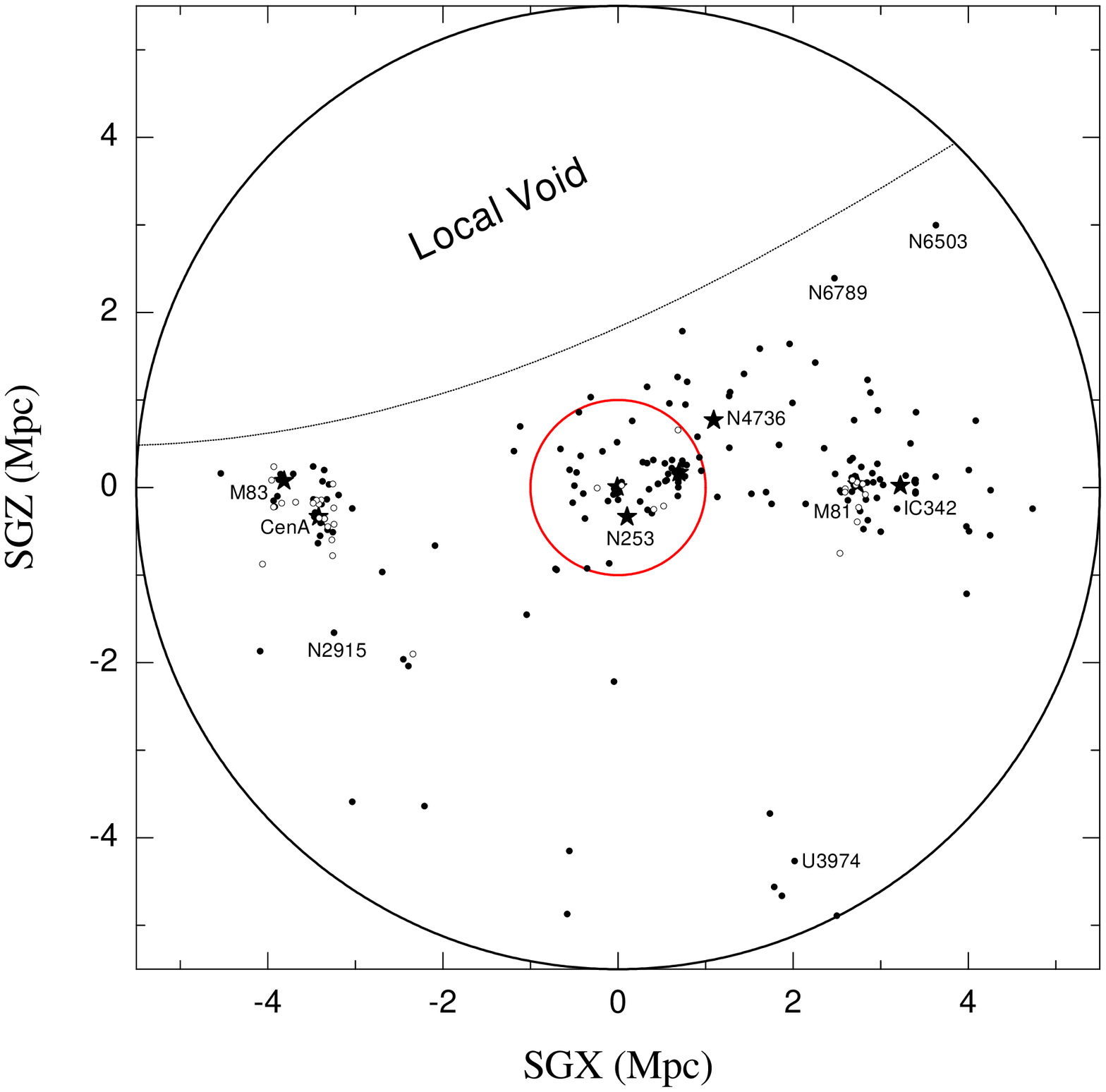}
\setcounter{figure}{6}
\vspace{-5mm}
\caption{continued}
\end{figure*}
\end{document}